\documentclass[twocolumn,aps,prl,showpacs, superscriptaddress, amsmath,amssymb]{revtex4}

\usepackage{graphicx}
\usepackage{dcolumn}
\usepackage{bm}
\usepackage{epstopdf}
\usepackage{color}
\usepackage{pdfpages}

\begin{document}

\title{Scaling of liquid-drop impact craters in wet granular media}
\author{Qianyun Zhang}
\affiliation{Department of Chemical Engineering and Materials
Science, University of Minnesota, Minneapolis, MN 55455}
\author{Ming Gao}
\affiliation{Department of Chemical Engineering and Materials
Science, University of Minnesota, Minneapolis, MN 55455}
\author{Runchen Zhao}
\affiliation{Department of Chemical Engineering and Materials
Science, University of Minnesota, Minneapolis, MN 55455}
\author{Xiang Cheng}
\affiliation{Department of Chemical Engineering and Materials
Science, University of Minnesota, Minneapolis, MN 55455}
\date{\today}

\begin{abstract}

Combining high-speed photography with laser profilometry, we study the dynamics and the morphology of liquid-drop impact cratering in wet granular media---a ubiquitous phenomenon relevant to many important geological, agricultural, and industrial processes. By systematically investigating important variables such as impact energy, the size of impinging drops and the degree of liquid saturation in granular beds, we uncover a novel scaling for the size of impact craters. We show that this scaling can be explained by considering the balance between the inertia of impinging drops and the strength of impacted surface. Such a theoretical understanding confirms that the unique energy partition originally proposed for liquid-drop impact cratering in dry granular media also applies for impact cratering in wet granular media. Moreover, we demonstrate that compressive stresses, instead of shear stresses, control granular impact cratering. Our study enriches the picture of generic granular impact cratering and sheds light on the familiar phenomena of raindrop impacts in granular media.                

\end{abstract}

\maketitle

\section{I. Introduction}

Walking on a beach after light rain, one may easily identify countless raindrop impact craters on the sand surface. Such a daily-life phenomenon is directly relevant to important geological and agricultural processes such as soil erosion \cite{Pimentel95,Furbish07}, drip irrigation \cite{Dasberg99} and dispersion of micro-organisms in soil \cite{Brodie51}. Understanding the dynamics of liquid-drop impact cratering in granular media may even help in revealing the properties of Earth's atmosphere in the geological past \cite{Som12} and the mechanism of asteroid impact cratering under extreme conditions \cite{Zhao15}. However, although solid-sphere impact cratering has been studied as early as the time of Robert Hooke \cite{Hooke65} and has already become one of the most extensively studied subjects in granular physics and fluid mechanics (\cite{Uehara03, Walsh03, Lohse04, Royer05, Katsuragi07, Clark14, Nordstrom14, Marston12, Takita13, Birch14, Suarez13} and references therein), liquid-drop impact cratering has only started to receive attention in recent years \cite{Katsuragi10, Marston10, Delon11, Katsuragi11, Nefzaoui12, Long14, Zhao15, Zhao15_1,Zhao15_2}. Moreover, current studies on liquid-drop impact cratering only focused on impact processes in dry granular media. Liquid-drop impact cratering in wet granular media has not been explored so far.   

Understanding liquid-drop impact cratering in wet granular media is practically more important. Under normal natural conditions, granular media such as soil always have non-zero water content, which can be quantified by the degree of saturation---a concept originated in soil science \cite{Jumikis84}. On the other hand, it has been shown that mechanical properties of granular media are considerably modified when mixed with even a small amount of liquid \cite{Hornbaker97, Nowak05, Scheel08}. Cohesive forces between granular particles induced by the surface tension of partially saturating liquid can dramatically increase the stiffness of granular media, which enables us to build sand castles, another fascinating phenomenon on the beach that manifests the unique properties of granular materials. As a result, we expect that the dynamics of liquid-drop impact cratering and the morphology of the resulting impact craters in wet granular media should differ sharply from their counterparts in dry granular media. Novel scalings of impact cratering are anticipated. 
 
Here, we experimentally study the dynamics and the morphology of liquid-drop impact cratering in wet granular media. Particularly, we seek the scaling relation between the size of impact craters and important control parameters of the problem including impact energy, the size of impinging drops, and the saturation of granular beds. Depending on the degree of saturation, the state of wet granular media can be categorized into four states in the order of increasing liquid content: pendular, funicular, capillary and slurry state \cite{Mitarai06}. In our study, we focus on the pendular state of granular media, where the liquid in granular beds exists only in the form of liquid bridges between the contact points of particles. A granular bed in the pendular state has low liquid content and is the most prevalent state of natural soil \cite{Jumikis84}. 
         
\begin{figure*}
\begin{center}
\includegraphics[width=7in]{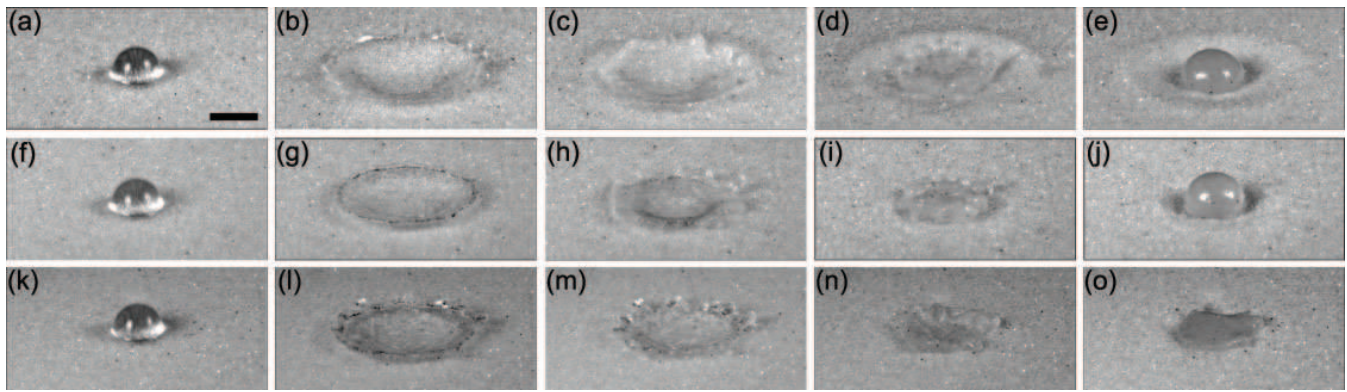}
\end{center}
\caption[impact dynamics]{Impact of an oil drop on a wet granular bed. Snapshots from high-speed movies showing the impact of a 2.6 mm oil drop with $E = 4.9\times10^{-5}$ J on granular beds with saturation $S = 0$ (a-e), $S = 0.3\%$ (f-j) and $S = 0.8\%$ (k-o). For $S=0$, the time elapsed after the initial impact is $t$ = 0.2 ms (a), 3.2 ms (b), 7.8 ms (c), 13.2 ms (d), and 85.7 ms (e). For $S=0.3\%$, $t$ = 0.2 ms (f), 2.0 ms (g), 7.4 ms (h), 12.1 ms (i), and 30.2 ms (j). For $S=0.8\%$, $t$ = 0.2 ms (k), 4.6 ms (l), 5.9 ms (m), 10.6 ms (n), and 37.9 ms (o). Scale bar: 2.6 mm.} \label{Figure1}
\end{figure*}

\section{II. Experiments}

In our experiments, a liquid drop of diameter $D_d$ is released from a height $h$ above the surface of a granular bed that has a fixed volume fraction of particles, $\phi = 0.59 \pm 0.01$, and a saturation of liquid, $S$. The granular bed is composed of soda-lime glass beads of diameter $d = 90\pm15$ $\mu$m. The saturation of the granular bed, $S$, is defined as $S = V_l/(V_l + V_a)$, where $V_l$ is the volume of liquid in the bed and $V_a$ is the volume of air void in the bed \cite{Marston12,Jumikis84}. $S = 0$ corresponds to a completely dry granular bed, while $S = 100\%$ corresponds to fully saturated bed. We limit $S$ below $1\%$ in our experiments in the pendular state of wet granular materials \cite{Mitarai06}, which is also classified as ``damp soil'' in soil mechanics \cite{Jumikis84}. At higher $S$, the yield stress of a granular bed becomes so large that the typical liquid drop we use in our experiments with $D_d = 2.6$ mm cannot create appreciable craters on the granular surface near its terminal velocity. 

To avoid the evaporation of liquid from the granular bed that will cause a time-dependent saturation during experiments, we choose light mineral oil (density $\rho_d = 0.84$ g/cm$^3$, viscosity $\eta = 28.7$ mPa$\cdot$s and surface tension $\sigma = 30$ mN/m) as our saturating liquid. Mineral oil has a vapor pressure smaller than 13 Pa at room temperature. Thus, the evaporation of the liquid is negligible during our experiments. For preparing a wet granular bed, we first dissolve a control amount of mineral oil in 200 ml of hexane. Then, 200 g of soda-lime glass beads are added into the mixture of mineral oil and hexane. The resulting slurry is sonicated for 30 minutes and then placed in a hood for 12 hours to allow for the evaporation of hexane. The process of sonication employs sound waves to agitate particles in the slurry, which promotes the mixing of mineral oil with particles. Finally, the granular particles are dried for 24 hours at 54 $^\circ$C to completely remove the excess hexane. The above procedure ensures that even a small amount of mineral oil can be uniformly coated on the surface of granular beads. We test the granular particles treated by hexane alone without mineral oil. The dynamics and the morphologies of liquid-drop impact cratering on the granular bed composed of such particles are indistinguishable from experiments on granular bed composed of original dry granular particles. Thus, hexane does not change the properties of our glass beads. 

We use the same mineral oil as our impinging drops, which have drop diameters of 1.8 mm $\le D_d \le$ 5.3 mm, covering the same size range of natural raindrops \cite{Villermaux09}. Although the impact energy of liquid drops, $E_{impact}$, can be approximated as $mgh$ for small $h$, the effect of air drag on liquid drops becomes significant at large $h$ such that $mgh$ is considerably larger than the true $E_{impact}$. Here, $g$ is the gravitational acceleration. Hence, instead of $mgh$, we calculate $E_{impact}$ based on $E_{impact} = \frac{1}{2}mU^2$, where $m$ is the mass of liquid drops and $U$ is the impact velocity of liquid drops immediately before the drops touch on the unperturbed surface of granular beds. $U$ is directly measured from high-speed photography.

\section{III. Dynamics and Morphology of Liquid-Drop Impact Cratering}

\begin{figure}
\begin{center}
\includegraphics[width=3.35in]{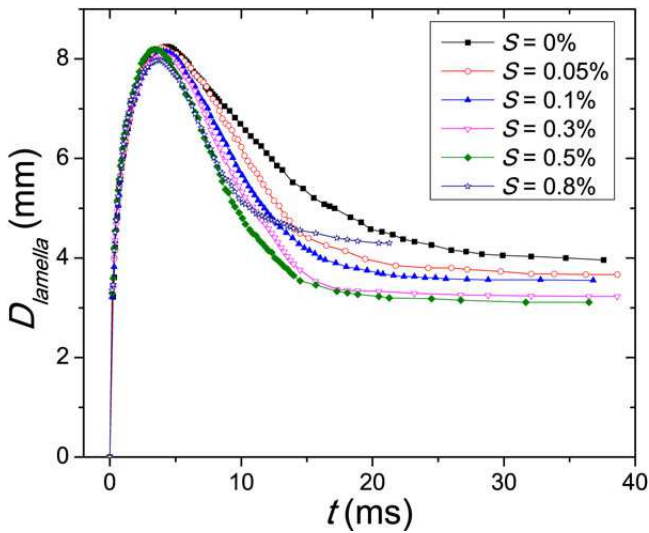}
\end{center}
\caption[Morphology of craters]{(Color online). Dynamics of liquid lamellae on granular bed of different liquid saturations. The diameter of liquid lamellae, $D_{lamella}$, as a function of time, $t$, is obtained from high-speed photography. The impacting liquid drops have a diameter $D_d = 2.6$ mm and an impact energy $E_{impact} = 4.9 \times 10^{-5}$ J. The saturations of the granular beds, $S$, are indicated in the plot. $t=0$ is defined as the time when the drops first touch the undisturbed surface of granular beds.} \label{Figure2}
\end{figure}

The process of impact cratering is recorded using a high-speed camera (Photron SA-X2) at the frame rate of 12,500 frames per second. Fig.~\ref{Figure1} shows the cratering dynamics for $D_d = 2.6$ mm drops impacting on granular bed of three different saturations at a fixed impact energy of $E_{impact} = 4.9\times10^{-5}$ J (see also the Supplemental Movies S1, S2 and S3). In all three cases, upon impact, the bottom of the drops first penetrates into the granular surface to create craters, while the upper part of the drops continuously deforms and spreads outward into liquid lamellae (Fig.~\ref{Figure1}a, f, k). Nevertheless, clear differences can be observed for beds with different saturations. For a dry granular bed, the penetration of impinging drop is deepest. The spreading lamella moves along the curved surface of crater that has the shape of a circular paraboloid (Fig.~\ref{Figure1}b). During the impact, a large amount of granular particles are ejected into the air (Movie S1). The lamella finally retracts after it reaches the maximum spreading diameter. During the retraction, the lamella entrains a layer of granular particles on its surface (Fig.~\ref{Figure1}c, d). The process comes to a halt with an approximately spherical liquid drop coated with a thin layer of granular particles---the so-called ``liquid marble'' \cite{Aussillous01, Zhao15}---sitting in the center of an appreciable crater (Fig.~\ref{Figure1}e). The liquid imbibes into the bed at long times on the order of seconds. The dynamics are qualitatively the same as that of impact catering of water drops on a dry granular bed \cite{Zhao15}.

\begin{figure}
\begin{center}
\includegraphics[width=3.35in]{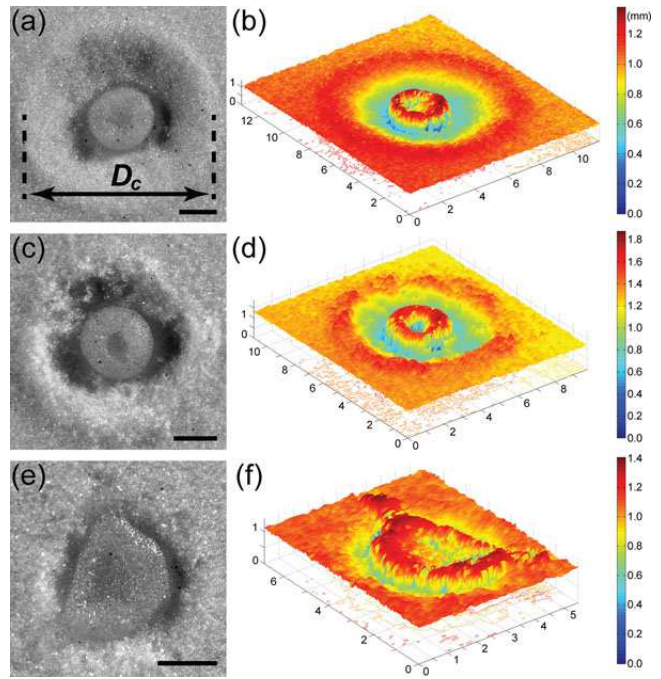}
\end{center}
\caption[Morphology of craters]{(Color online). Morphology of impact craters resulted from the impact of oil drops with drop diameter $D_d = 2.6$ mm and impact energy $E_{impact} = 4.9 \times 10^{-5}$ J. Granular beds have saturation $S = 0$ (a, b), $S = 0.3\%$ (c, d), and $S=0.8\%$ (e, f). (a), (c) and (e) are from optical photography. Scale bars: 2.0 mm. (b), (d) and (f) are the 3D topography of the corresponding craters measured from laser profilometry. The colors indicate the height.} \label{Figure3}
\end{figure}

At saturation $S = 0.3\%$, liquid bridges form between the contact points of granular particles, which induce cohesive forces among particles \cite{Mitarai06}. Consequently, the number of particles that are ejected into the air is reduced (Movie S2). The impact ends up creating a smaller and less symmetric crater compared with the crater formed in the dry granular bed (Fig.~\ref{Figure1}j and Fig.~\ref{Figure3}c). Since the liquid lamella retracts back to a liquid marble, the granular residue in the center of the crater maintains its symmetric spherical shape. The change of dynamics becomes more obvious at even higher saturation of $S = 0.8\%$. At this relatively high saturation, the liquid lamella spreads outward in a shallow angle on the granular surface (Fig.~\ref{Figure1}l). Very few ejected particles are observed (Movie S3). Furthermore, due to the strong cohesion between particles, the retracting lamella experiences much larger resistance when lifting up and entraining surface particles. As a result, the lamella cannot fully retract back to a spherical drop and stops as a liquid puddle in the center of a barely recognizable crater. The edge of the puddle is pinned on the granular surface (Fig.~\ref{Figure1}o). Therefore, the granular residue formed at the end of this process is much less symmetric than the liquid marbles formed at low saturations. Due to the coating of liquid film on granular particles, the imbibition of liquid is faster for a high-saturation bed. Finally, for saturations larger than $1\%$, the dynamics of liquid-drop impacts are qualitatively similar to the dynamics of drop impacts on rough solid surfaces \cite{Xu07}, with the exception of fast liquid imbibition toward the end of impact processes.   

Quantitatively, the dynamics of the spreading and retracting of liquid lamellae on granular beds of different saturations are shown in Fig.~\ref{Figure2}. The diameter of liquid lamellae, $D_{lammella}$, shows a non-monotonic trend, representing the spreading process before the peak and the retracting process after the peak. The maximal diameter of liquid lamellae during impact is independent or only weakly depends on the saturation of granular beds. The final size of liquid marbles after retracting decreases when $S$ increases from 0 to $0.5\%$. However, for the highest saturation we study in our experiments at $S=0.8\%$, the retracting stops before the liquid marble can form. As a result, the size of the final liquid puddle is larger than that of liquid marbles formed at lower $S$.    

\begin{figure}
\begin{center}
\includegraphics[width=3.35in]{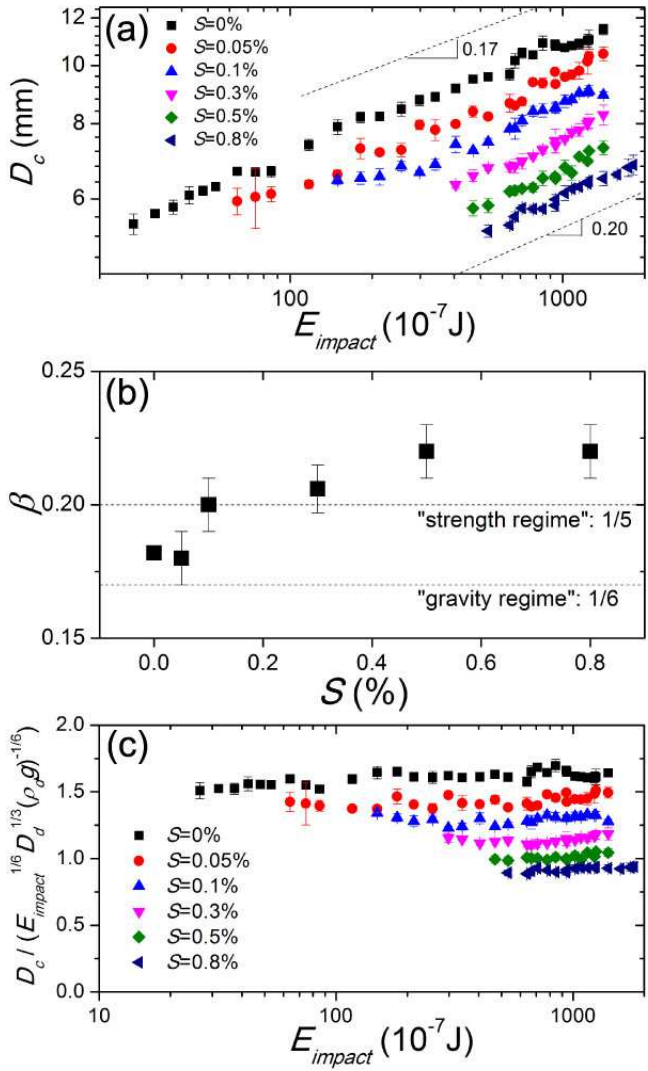}
\end{center}
\caption[Crater diameters]{(Color online). Impact-energy dependence of the size of impact craters. (a) Crater diameter, $D_c$, as a function of impact energy, $E_{impact}$, for granular bed of different saturations. The diameter of oil drops $D_d = 2.6$ mm. Dashed lines indicate the slope of 0.17 and 0.20, respectively. The lowest impact energy, at which we can observe appreciable craters, increases with $S$.  (b) The power-law exponent, $\beta$, in the scaling $D_c \sim E_{impact}^\beta$ as a function of bed saturation, $S$. The exponents predicted from the scaling in the strength regime and in the gravity regime are indicated by the dashed lines. (c) Scaled $D_c$ as a function of $E_{impact}$ based on the Schmidt-Holsapple scaling (Eq.~\ref{equ1}).} \label{Figure4}
\end{figure}

Fig.~\ref{Figure3} shows the morphology of impact craters at the end of impact processes after the liquid drains into the bed. The three-dimensional (3D) surface topography of craters is measured using a laser profilometer (Kenyence LJ-V7060) (Fig.~\ref{Figure3}b, d, f), which allows us to accurately measure the rim-to-rim diameter of craters, $D_c$ (Fig.~\ref{Figure3}a). For less symmetric craters at large $S$, the rim-to-rim diameter of a crater is obtained by averaging the diameters measured along two orthogonal directions through the geometric center of the crater. The horizontal direction is chosen along the laser scanning line of the laser profilometer, which is randomly orientated with respect to the crater. To further reduce statistical errors, at least three different impact craters are measured for each set of control parameters. 

$D_c$ as a function of impact energy, $E_{impact}$, for beds with different saturations is shown in Fig.~\ref{Figure4}a. We fit our data using a power-law scaling, $D_c \sim E_{impact}^\beta$ \cite{Note15}. For granular beds of low saturations, the power exponent is $\beta = 0.18 \pm 0.01$, consistent with the previous studies on liquid-drop impacts in dry granular media \cite{Zhao15, Nefzaoui12}. However, for granular beds with higher saturations, the exponent increases to a larger value $\beta = 0.22 \pm 0.01$. We plot the power-law exponents for each saturation in Fig.~\ref{Figure4}b, which shows a clear trend of the increase of $\beta$ with saturations.

\section{IV. Scaling of liquid-drop impact craters}
The 0.18 power-law scaling of dry and low saturation granular beds can be quantitatively understood as the Schmidt-Holsapple (S-H) scaling originally proposed for asteroid impact cratering \cite{Zhao15, Holsapple93}:
\begin{eqnarray}
\label{equ1}  
D_c & \simeq & (\rho_d g)^{-1/6}\cdot D_d^{1/3}\cdot E_{impact}^{1/6} \\
    & \simeq & g^{-1/6} \cdot D_d^{5/6} \cdot U^{1/3} \nonumber
\end{eqnarray} 
(see Eq. 22b in Ref.~\cite{Holsapple93}). During liquid-drop impact cratering, only part of the impact energy is converted to the kinetic energy of ejected particles, which creates impact craters against gravitational potential of particles. Zhao et al. proposed that the fraction of impact energy that is converted to the kinetic energy of particles is given by $f = (D_d/D_c)^2$, which leads to the S-H scaling when $f \cdot E_{impact}$ is balanced by the gravitational potential of particles \cite{Zhao15}. $f$ is the coefficient characterizing the energy partition in liquid-drop impact cratering processes. However, when we directly apply the S-H scaling to $D_c$ at different $S$, our data do not collapse into a single master curve (Fig.~\ref{Figure4}c).

To explain the increase of power-law exponents with $S$ shown in Fig.~\ref{Figure4}b, we need to explicitly consider the effect of the yield stress of wet granular beds. It is instructive to first review the scaling of solid-sphere impact cratering on a material with non-zero yield stress $Y$ \cite{Holsapple93}. Grouping all relevant parameters, we have four dimensionless numbers in the impact-cratering problem, which follow the general relation: 
\begin{equation}
\label{equ2} {\frac{\rho_d V}{m}=F\left(\frac{gD_d}{U^2},\frac{Y}{\rho_dU^2},\frac{\rho_d}{\rho}\right)}.
\end{equation}
Here, $V$ is the volume of crater and $m \equiv \pi/6\rho_dD_d^3$ and $U$ are the mass and the impact velocity of the projectile, respectively. $\rho$ is the density of the impacted surface. $F(x)$ is some unknown function. In our case, since $\rho_d/\rho \simeq 1$ is kept constant, the third term in $F$ is not directly relevant. Two limiting regimes can be deduced from Eq.~\ref{equ2} \cite{Note15}: ({\it i}) the ``gravity regime'', $\rho_d V/m = F(gD_d/U^2)$, where the strength of material is much smaller than the gravitational pressure; and ({\it ii}) the ``strength regime'',  $\rho_d V/m =F(Y/\rho_dU^2)$, where the strength is much larger than the gravitational pressure. To obtain the final scaling in both regimes, we further assume that the dependence of $V$ on $D_d$ and $U$ is through a specific combination $D_d^3U^2 \sim mU^2 \simeq E_{impact}$, i.e., the impact energy of projectile. This assumption can be seen as a special case of point-source solutions \cite{Holsapple82,Holsapple87}. Applying the assumption in the gravity regime, we reach $\rho_d V/m \simeq (gD_d/U^2)^{-3/4}$, which leads to $V \simeq (E_{impact}/\rho_dg)^{3/4}$. Since $V \simeq D_c^3$, we have $D_c \sim E_{impact}^{1/4}$. The scaling successfully explains experimental observations on solid-sphere impacting cratering on dry granular beds \cite{Uehara03, Walsh03}, where the yield stress of a granular bed is small compared with the gravitational pressure. By contrast, in the strength regime, the same assumption leads to $\rho_d V/m \simeq (Y/\rho_d U^2)^{-1}$, which gives $V \simeq E_{impact}/Y$ and $D_c \sim E_{impact}^{1/3}$.

\begin{figure}
\begin{center}
\includegraphics[width=3.35in]{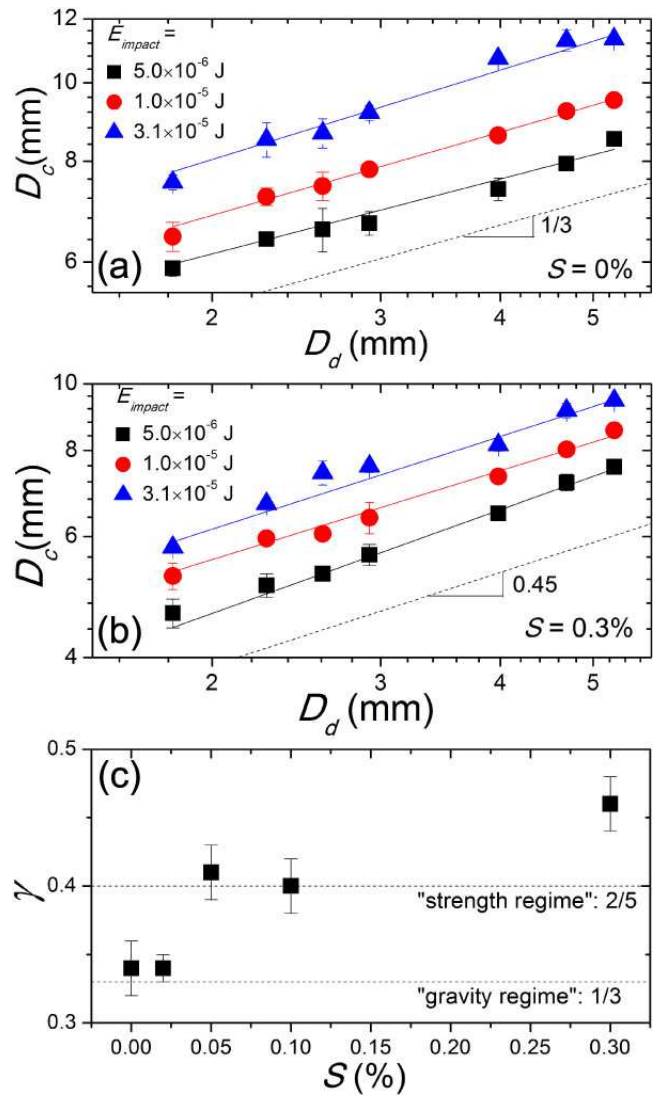}
\end{center}
\caption[Drop size dependence]{(Color online). Drop-size dependence of the size of impact craters. $D_c$ as a function of the diameter of liquid drops, $D_d$, at three impact energies for granular beds with $S=0$ (a) and $S=0.3\%$ (b). Solid lines are power-law fittings. Dashed lines indicate the power exponents 1/3 and 0.45, respectively. (c) The power-law exponent, $\gamma$, in the scaling $D_c \sim D_d^\gamma$ as a function of bed saturation, $S$. Data are averaged over all three $E_{impact}$. The exponents predicted from the scaling in the strength regime and in the gravity regime are indicated by the dashed lines.} \label{Figure5}
\end{figure}
       
We will extend the above scaling argument for solid-sphere impact cratering to liquid-drop impact cratering. First, following a previous study on liquid-drop impact cratering in dry granular media \cite{Zhao15}, we assume that the fraction of impact energy converted for creating impact craters follows the same formula, i.e., $f = (D_d /D_c)^2$. The rest of impact energy $(1-f) \cdot E_{impact}$ turns into the dissipation and surface energy of spreading lamellae. However, for wet granular beds of high saturations, the yield stress should be larger than the gravitational pressure. In other words, the cratering occurs in the strength regime. Hence, differing from the case of dry granular media, the effective impact energy for cratering, $f \cdot E_{impact}$, is mainly used to overcome the yield stress of wet granular media instead of the gravitational pressure of particles. While applying $f$ in the gravity regime leads to the S-H scaling \cite{Zhao15}, when applying the same $f$ in the strength regime, we achieve
\begin{equation}
\label{equ3} {V=C_1\frac{f\cdot E_{impact}}{Y}},
\end{equation} 
where $C_1$ is a numerical constant. Assume that a liquid-drop impact crater has the shape of a circular paraboloid. Then, $V = \alpha \pi D_c^3/8$. Here, $\alpha$ is the ratio of the depth to the diameter of the crater, which we assume is a constant independent of $S$ and $E_{impact}$. It is difficult, if not impossible, to measure the depth of craters underneath the mixture of liquid and granular particles at the center of impact craters. The assumption of a constant $\alpha$ is inspired by our previous measurements of the depth of craters for liquid-drop impact cratering in dry granular media, where a liquid drop jumps off the surface of the granular bed during impact that allows for direct imaging of the bottom of impact craters \cite{Zhao15}. Moreover, such an assumption has also been widely used for understanding the scaling relation for solid-sphere impact cratering \cite{Uehara03, Walsh03, Nordstrom14, Suarez13} and has been shown to be valid in a wide range of impact energies \cite{Walsh03}. With a constant $\alpha$, we insert $V$ into Eq.~\ref{equ3} to achieve 
\begin{equation}
\label{equ4} {D_c=C_2 \cdot D_d^{2/5}\cdot E_{impact}^{1/5}\cdot Y^{-1/5}},
\end{equation} 
where $C_2 = (8C_1/\alpha\pi)^{1/5}$ is a dimensionless constant. This scaling result shows that when transitioning from the gravity regime to the strength regime, the power-law exponent, $\beta$, in the scaling of $D_c \sim E_{impact}^{\beta}$ should increase from 1/6 of the S-H scaling (Eq.~\ref{equ1}) to 1/5 (Eq.~\ref{equ4}), qualitatively agreeing with our experimental observation (Fig.~\ref{Figure4}b). 

Furthermore, Eq.~\ref{equ4} also predicts that the dependence of $D_c$ on the size of impinging drops, $D_d$, should change from a 1/3 scaling for a dry granular bed (Eq.~\ref{equ1}) to a 2/5 scaling for a wet granular bed. Fig.~\ref{Figure5}a and b show our measurements on $D_c$ as a function of $D_d$ for different saturations. The results are fitted with $D_c \sim D_d^\gamma$. Although the dynamic range of $D_d$ is limited due to the size range of liquid drops one can normally obtain in laboratory conditions \cite{Villermaux09}, $\gamma$ indeed increases from 0.33 to 0.45, consistent with the prediction (Fig.~\ref{Figure5}c).

\begin{figure}
\begin{center}
\includegraphics[width=3.35in]{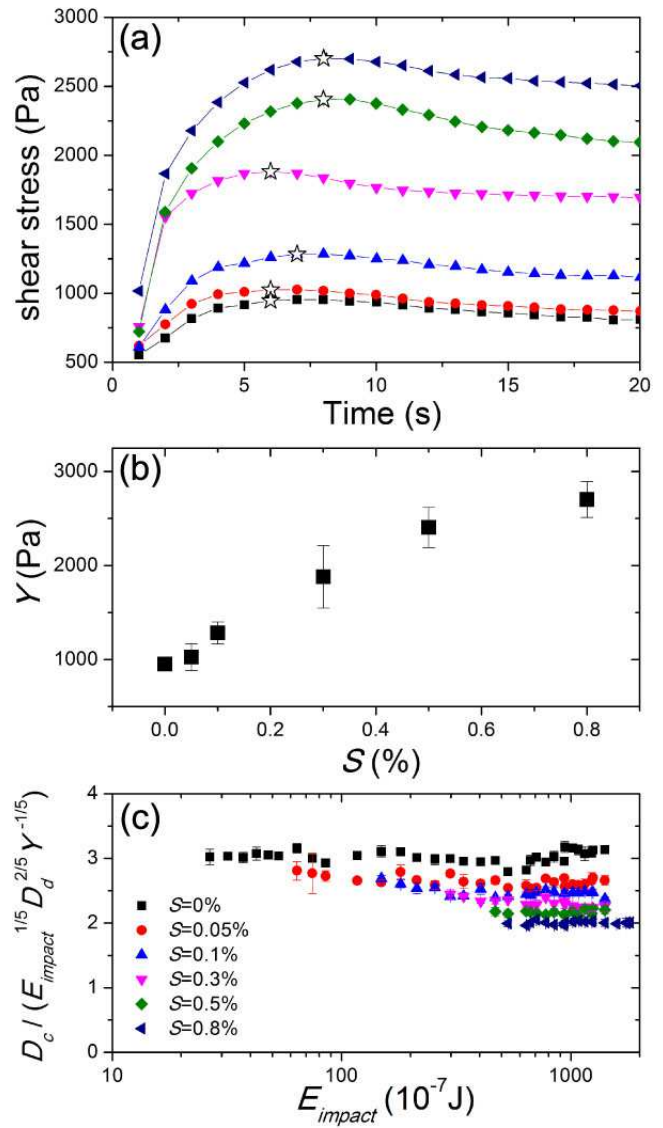}
\end{center}
\caption[Vane-yield stress dependence]{(Color online). Shear stress dependence of the size of impact craters. (a) Shear stress measured from a rotating vane rheometer versus time. Time here is equivalent to the shear strain applied by the rheometer, since a constant shear rate of 0.01 rad/s is applied. From bottom to top, the saturation of the granular bed is $S = 0$, $0.05\%$, $0.1\%$, $0.3\%$, $0.5\%$, and $0.8\%$. The peak of the stresses (stars) defines the shear yield stress of a granular bed, $Y$. (b) Shear yield stress from rotating vane rheometry, $Y$ as a function of $S$. (c) Scaled $D_c$ as a function of $E_{impact}$ based on the strength regime scaling (Eq.~\ref{equ4}), where $Y$ are measured from rotating vane rheometry.} \label{Figure6}
\end{figure}
       
Last, to verify the full scaling of $D_c$ predicted by Eq.~\ref{equ4}, we need to quantify the strength of a wet granular bed, $Y$. Two different methods are adopted to measure the yield stress of beds at different saturations. Firstly, following previous studies on solid-sphere impact cratering in wet granular media \cite{Takita13}, we use a rotating vane rheometer (TA AR-G2) to measure the shear stress of wet beds (Fig.~\ref{Figure6}a). The vane spindle has a diameter of 28 mm and a height of 42 mm and is rotated at 0.01 rad/s in the quasi-static limit. The yield stress can be defined as the peak in the shear stress-time curve (Fig.~\ref{Figure6}a), which increases with saturation as expected (Fig.~\ref{Figure6}b). However, when we replace $Y$ in Eq.~\ref{equ4} with the yield stress thus obtained, the scaled $D_c$ do not collapse  (Fig.~\ref{Figure6}c). The result indicates that the yield stress measured from rotating vane rheometry may not be the relevant parameter to quantify the strength of a granular bed in the context of impact cratering.

\begin{figure}
\begin{center}
\includegraphics[width=3.35in]{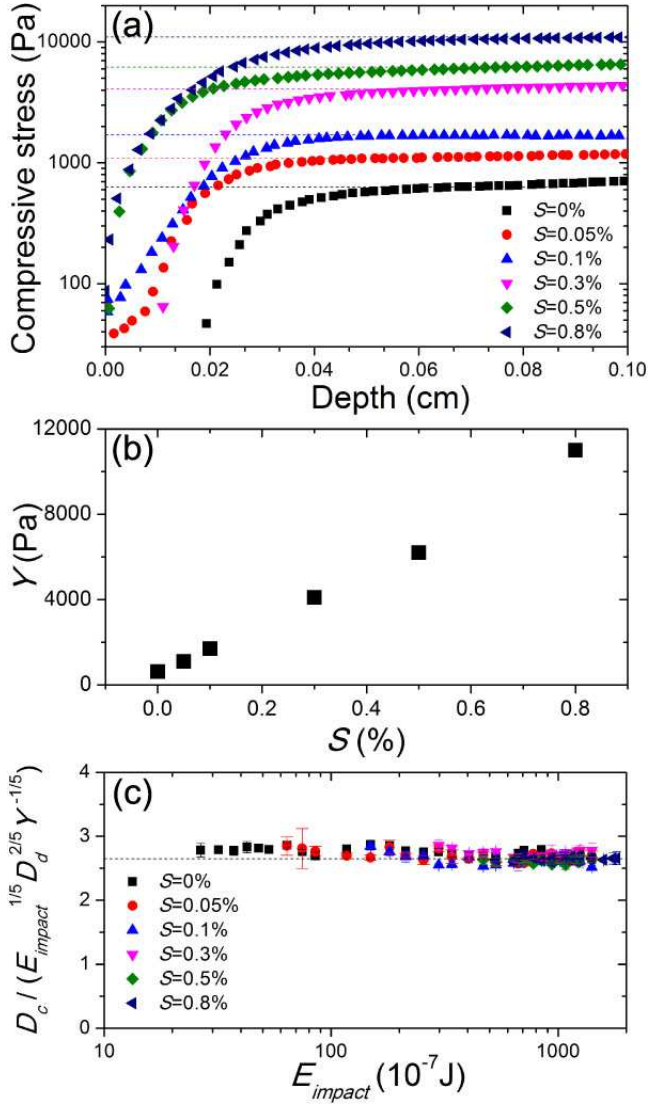}
\end{center}
\caption[Compressive-yield stress dependence]{(Color online). Compressive stress dependence of the size of impact craters. (a) Compressive stress versus penetration depth. The plateau of stresses (dashed lines) defines the compressive yield stress of a granular bed, $Y$. (b) Compressive yield stress from compression experiments, $Y$, as a function of $S$. (c) Scaled $D_c$ as a function of $E_{impact}$ based on the strength regime scaling (Eq.~\ref{equ4}), where $Y$ are measured from compression experiments.} \label{Figure7}
\end{figure}
 
When a solid sphere or a liquid drop impacts on a granular surface, granular particles on the flat surface yield mainly due to compression rather than shear. Therefore, the yield stress of a granular bed under compression is more appropriate in characterizing the strength of a wet granular bed under impact. Thus, in the second method, we measure the compressive stress of a granular bed by compressing an 8 mm-in-diameter stainless steel circular plate into a wet granular bed at a rate of 0.1 mm/s in the quasi-static limit. The normal force or stress on the plate is recorded by using a rheometer equipped with normal force gauges (TA G2-RSA). For a typical stress-distance curve (Fig.~\ref{Figure7}a), the compressive stress first increases sharply when the plate touches the surface of a granular bed. Granular particles rearrange under the plate to adapt for the flat surface of the plate. The bed underneath the plate is compressed slightly with the average spacing between the granular particles reduced. When the plate is further pushed into the media, the bed yields. Particles are pushed upward against gravity around the circumference of the plate. In this regime, the stress exhibits a plateau, which can be used to quantify the yield stress of a granular bed under compression (Fig.~\ref{Figure7}a). When the plate is pushed even further into the media beyond the plateau regime, the compressive stress increases again in an approximately linear fashion (not shown) \cite{stone04}. Since the depth of liquid-drop impact craters are typically a couple of millimeters (Fig.~\ref{Figure3}b, d, f), the yield stress measured near the surface of a granular bed is directly relevant to liquid-drop impact cratering. Compressive yield stresses are larger than shear yield stresses obtained from rotating vane rheometry (Fig.~\ref{Figure7}b). Replacing $Y$ in Eq.~\ref{equ4} with the yield stress from compression measurements leads to an excellent collapse of all our experimental data (Fig.~\ref{Figure7}c). The numerical prefactor in Eq.~\ref{equ4}, $C_2 = 2.65$, on the order of one. 

\begin{figure}
\begin{center}
\includegraphics[width=3.35in]{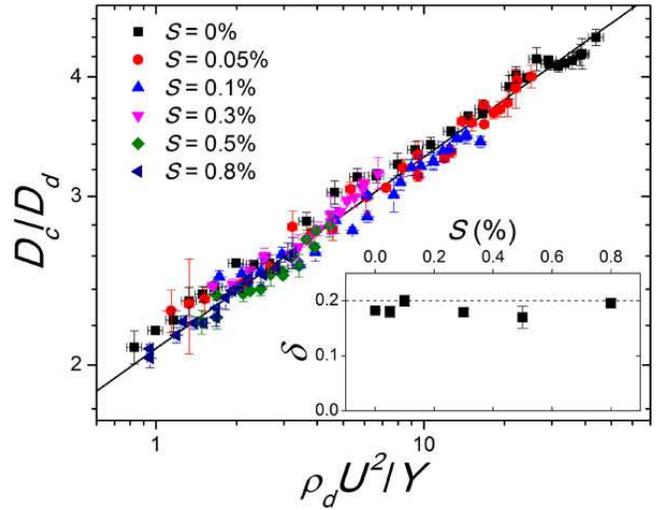}
\end{center}
\caption[Scaling of craters]{(Color online). Scaling of liquid-drop impact craters. Dimensionless crater diameter, $D_c/D_d$, as a function of dimensionless inertia, $\rho_dU^2/Y$, for granular beds of different saturations. Craters are generated by the impact of oil drops with $D_d = 2.6$ mm. The solid line has a slope of 0.20. Inset: The power-law exponent, $\delta$, in the scaling $D_c/D_d \simeq \left(\rho_dU^2/Y\right)^\delta$ as a function of bed saturation, $S$. The dashed line indicates the exponent predicted from the strength regime scaling.} \label{Figure8}
\end{figure}

The scaling in Eq.~\ref{equ4} can also be presented in terms of the dimensionless crater diameter, $D_c/D_d$. According to Eq.~\ref{equ4}, $D_c/D_d$ should follow a simple power-law scaling with the dimensionless inertia, $\rho_dU^2/Y$: $D_c/D_d \simeq (\rho_dU^2/Y)^\delta$ with $\delta = 1/5$. We plot $D_c/D_d$ as a function of $\rho_dU^2/Y$ in Fig.~\ref{Figure8}. As expected, all the data collapse into a master curve. Power-law fittings at different $S$ yield $\delta = 0.19 \pm 0.01$ (Fig.~\ref{Figure8} inset), quantitatively agreeing with our prediction.  

\section{V. Conclusion}
We studied the dynamics of liquid-drop impact cratering in wet granular media and systematically investigated the dependence of the morphology of resulting impact craters on the saturation of granular beds, impact energy and the size of impinging drops. A novel scaling of the size of impact craters has been revealed. We further showed that such a scaling can be quantitatively understood by combining the scaling of solid-sphere impact cratering in the strength regime with the unique energy partition of liquid-drop impact cratering. In particular, we showed that the coefficient characterizing the energy partition of liquid-drop impacts follows a simple formula, $f = (D_d /D_c)^2$, independent of the saturation of granular beds. Moreover, we demonstrated that the strength of a granular bed relevant to impact cratering should be measured through compression rather than shear experiments. 

The results from our study are directly relevant to the geological processes of raindrop impact cratering and should be useful in interpreting the morphologies of raindrop impact craters. For example, the size of raindrop impact craters on fossilized granular beds have been used to infer the properties of Earth's atmosphere in the geological past \cite{Som12,Metz81}. Som and co-workers estimated the air density on Earth 2.7 billion years ago from the terminal velocity of raindrops, where the relation between the size of impact craters and the terminal velocity of water drops is critical for the accuracy of the estimate \cite{Som12}. Here, we demonstrate that the dynamics and the morphologies of raindrop impact cratering in wet granular media qualitatively differ from those in dry granular media even when the bed contains only a small amount of liquid, a condition that is almost unavoidable in raining. Thus, the lower limit of the terminal velocity of raindrops and, therefore, the upper limit of the air density on Earth estimated in Ref.~\cite{Som12} need to be revisited to explicitly consider the degree of liquid saturations.                

Finally, although increasing $S$ above a few percentages into the funicular state of granular media will result in a granular bed that is too stiff to be able to produce any appreciable liquid-drop impact craters, increasing $S$ further into the capillary and slurry state can reverse the trend and considerably reduce the yield stress of a granular bed \cite{Hornbaker97, Nowak05, Scheel08, Mitarai06}. Thus, we expect that craters can be created by the impacts of liquid drops on granular beds with very high liquid content. Such processes are mainly responsible for impact craters observed on fully saturated soil after heavy storms. Liquid-drop impact cratering on granular beds with high $S$ will be the subject of our future research.

We thank D. Giles, F. Jarpardi, W. Teddy, H. Tjugito and J. Wang for help with experiments. The research is supported by NSF CAREER DMR-1452180. Q. Z. and R. Z. also acknowledge the support by the undergraduate research opportunities program from University of Minnesota.


\begin{thebibliography}{99}

\bibitem{Pimentel95}	D. Pimentel et al., Science {\bf 267}, 1117 (1995).
\bibitem{Furbish07} D. J. Furbish, K. K. Hamner, M. Schmeeckle, M. N. Borosund, and S. M. Mudd, J. Geophys. Res.-Earth {\bf 112}, F01001 (2007).
\bibitem{Dasberg99}	S. Dasberg, and D. Or, {\it Drip Irrigation} (Springer-Verlag, New York, 1999).
\bibitem{Brodie51}	H. J. Brodie, Can. J. Botany {\bf 29}, 224 (1951).
\bibitem{Som12}	S. M. Som et al., Nature {\bf 484}, 359 (2012).
\bibitem{Zhao15} R. Zhao, Q. Zhang, H. Tjugito, and X. Cheng, Proc. Natl. Acad. Sci. USA {\bf 112}, 342 (2015). 
\bibitem{Hooke65} R. Hooke, {\it Micrographia} (J. Martyn and J. Allestry, London, UK, 1665). 
\bibitem{Uehara03}	J. S. Uehara, M. A. Ambroso, R. P. Ojha, and D. J. Durian, Phys. Rev. Lett. {\bf 90}, 194301 (2003).
\bibitem{Walsh03}	A. M. Walsh, K. E. Holloway, P. Habdas, and J. R. de Bruyn, Phys. Rev. Lett. {\bf 91}, 104301 (2003).
\bibitem{Lohse04}	D. Lohse, R. Rauhe, R. Bergmann, and D. van der Meer, Nature {\bf 432}, 689 (2004).
\bibitem{Royer05}	J. R. Royer et al., Nat. Phys. {\bf 1}, 164 (2005).
\bibitem{Katsuragi07}	H. Katsuragi, and D. J. Durian, Nat. Phys. {\bf 3}, 420 (2007).
\bibitem{Clark14}	A. H. Clark, A. J. Petersen, and R. P. Behringer, Phys. Rev. E {\bf 89}, 012201 (2014).
\bibitem{Nordstrom14} K. N. Nordstrom, E. Lim, M. Harrington, and W. Losert, Phys. Rev. Lett. {\bf 112}, 228002 (2014).
\bibitem {Marston12} J. O. Marston, I. U. Vakarelski, and S. T. Thoroddsen, Phys. Rev. E {\bf 86}, 020301(R) (2012).
\bibitem{Takita13} H. Takita and I. Sumita, Phys. Rev. E {\bf 88}, 022203 (2013).
\bibitem{Birch14} S. P. D. Birch, M. Manga, B. Delbridge, and M. Chamberlain, Phys. Rev E {\bf 90}, 032208 (2014).
\bibitem{Suarez13}	J. C. Ruiz-Su{\'a}rez, Rep. Prog. Phys. {\bf 76}, 066601 (2013).


\bibitem{Katsuragi10}	H. Katsuragi, Phys. Rev. Lett. {\bf 104}, 218001 (2010).
\bibitem{Marston10}	J. O. Marston, S. T. Thoroddsen, W. K. Ng, and R. B. H. Tan, Powder Technol. {\bf 203}, 223 (2010).
\bibitem{Delon11}	G. Delon, D. Terwagne, S. Dorbolo, N. Vandewalle, and H. Caps, Phys. Rev. E {\bf 84}, 046320 (2011).
\bibitem{Katsuragi11}    H. Katsuragi, J. Fluid Mech. {\bf 675}, 552 (2011).
\bibitem{Nefzaoui12}	E. Nefzaoui, and O. Skurtys, Exp. Therm. Fluid Sci. {\bf 41}, 43 (2012).
\bibitem{Long14}	E. J. Long et al., Phys. Rev. E {\bf 89}, 032201 (2014).
\bibitem{Zhao15_1} S.-C. Zhao, R. de Jong, and D. van der Meer, Soft Matter {\bf 11}, 6562 (2015).
\bibitem{Zhao15_2} R. Zhao, Q. Zhang, H. Tjugito, and X. Cheng, Phys. Fluids {\bf 27}, 091111 (2015). 

\bibitem{Jumikis84} A. R. Jumikis, {\it Soil Mechanics} (Krieger Publishing Company, Malabar FL, 1984).
\bibitem{Hornbaker97} D. J. Hornbaker, R. Albert, I. Albert, A.-L. Barab{\'a}si, and P. Schiffer, Nature {\bf 387}, 765 (1997).
\bibitem{Nowak05} S. Nowak, A. Samadani, and A. Kudrolli, Nature Phys. {\bf 1}, 50 (2005).
\bibitem{Scheel08} M. Scheel, R. Seemann, M. Brinkmann, M. Di Michiel, A. Sheppard, B. Breidenbach, and S. Herminghaus, Nature Mater. {\bf 7}, 189 (2008). 
\bibitem{Mitarai06} N. Mitarai and F. Nori, Adv. Phys. {\bf 55}, 1 (2006).  

\bibitem{Villermaux09} E. Villermaux and B. Bossa, Nature Phys. {\bf 5}, 697 (2009).
\bibitem{Aussillous01} P. Aussillous and D. Qu{\'e}r{\'e}, Nature {\bf 411}, 924 (2001).
\bibitem{Xu07} L. Xu, Phys. Rev. E {\bf 75} 056316 (2007).

\bibitem{Note15} In the paper, we use the ``$=$'' sign to indicate the exact equality of two quantities including all numerical coefficients. ``$\simeq$'' is used when we ignore numerical coefficients but keeping all dimensional factors. ``$\sim$'' is used when we only express the power law relation between two quantities that do not have the same dimension.

\bibitem{Holsapple93} K. A. Holsapple, Annu. Rev. Earth Pl. Sc. {\bf 21}, 333 (1993). 
\bibitem{Holsapple82}	K. A. Holsapple, and R. M. Schmidt, J. Geophys. Res. {\bf 87}, 1849 (1982).
\bibitem{Holsapple87}	K. A. Holsapple, and R. M. Schmidt, J. Geophys. Res. {\bf 92}, 6350 (1987).
\bibitem{stone04} M. B. Stone, D. P. Bernstein, R. Barry, M. D. Pelc, Y.-K. Tsui, and P. Schiffer, Nature {\bf 427}, 503 (2004). 
\bibitem{Metz81} R. Metz, J. Sedim. Petrol. {\bf 51}, 265 (1981).



\end{thebibliography}
\end{document}